\documentclass[a4paper,fleqn,usenatbib]{mnras}

\usepackage{mathptmx}

\usepackage[T1]{fontenc}
\usepackage{ae,aecompl}
\usepackage{graphicx}   
\usepackage{graphics}
\usepackage{amsmath}    
\usepackage{amssymb}    
\usepackage[normalem]{ulem}

\title
[Accretion-disc precession in UX Ursae Majoris]
{Accretion-disc precession in UX Ursae Majoris}

\author[E. de Miguel et al.]
{
E. de Miguel,$^{1,2}$
\thanks{E-mail: \href{mailto:edmiguel63@gmail.com}{edmiguel63@gmail.com}}
J. Patterson,$^{3}$
D. Cejudo,$^{4}$
J. Ulowetz,$^{5}$
J. L. Jones,$^{6}$
\newauthor
J. Boardman,$^{7}$
D. Barret,$^{8}$
R. Koff,$^{9}$
W. Stein,$^{10}$
T. Campbell,$^{11}$
\newauthor
T. Vanmunster,$^{12}$
K. Menzies,$^{13}$
D. Slauson,$^{14}$
W. Goff,$^{15}$
G. Roberts,$^{16}$
\newauthor
E. Morelle,$^{17}$
S. Dvorak,$^{18}$
F.-J. Hambsch,$^{19}$
D. Starkey,$^{20}$
D. Collins,$^{21}$
\newauthor
M. Costello,$^{22}$
M. J. Cook,$^{23}$
A. Oksanen,$^{24}$
D. Lemay,$^{25}$
L. M. Cook,$^{26}$
\newauthor
Y. Ogmen,$^{27}$
M. Richmond,$^{28}$
and J. Kemp$^{29}$
\newauthor
\\
$^{1}$ Departamento de F\'{\i}sica Aplicada, Facultad de Ciencias
Experimentales, Universidad de Huelva, 21071 Huelva, Spain\\
$^{2}$ CBA-Huelva, Observatorio del CIECEM, Parque
Dunar, Matalasca\~nas, 21760 Almonte, Huelva, Spain\\
$^{3}$ Department of Astronomy, Columbia University, 550 West 120th Street,
New York, NY 10027, USA\\
$^{4}$ CBA-Madrid, Camino de las Canteras 42, Buz\'on 5, La Pradera del Amor,
El Berrueco, 28192 Madrid, Spain\\
$^{5}$ CBA-Illinois, Northbrook Meadow Observatory, 855 Fair Lane, Northbrook, 
IL 60062, USA\\
$^{6}$ CBA-Oregon, Jack Jones Observatory, 22665 Bents Road NE, Aurora, OR, USA\\
$^{7}$ CBA-Wisconsin, Luckydog Observatory, 65027 Howath Road, de Soto, WI 54624, USA\\
$^{8}$ CBA-France, 6 Le Marouzeau, St Leger Bridereix, 2300, France\\
$^{9}$ CBA-Colorado, Antelope Hills Observatory, 980 Antelope Drive West,
Bennett, CO 80102, USA\\
$^{10}$ CBA-Las Cruces, 6025 Calle Paraiso, Las Cruces, NM 88012, USA\\
$^{11}$ CBA-Arkansas, 7021 Whispering Pine Road, Harrison, AR 72601, USA\\
$^{12}$ CBA-Belgium, Walhostraat 1A, B-3401 Landen, Belgium\\
$^{13}$ CBA-Massachusetts, 318A Potter Road, Framingham, MA 01701, USA\\
$^{14}$ CBA-Iowa, Owl Ridge Observatory, 73 Summit Avenue NE, Swisher,
IA 52338, USA\\
$^{15}$ CBA-California, 13508 Monitor Lane, Sutter Creek, CA 95685, USA\\
$^{16}$ CBA-Tennessee, 2007 Cedarmont Drive, Franklin, TN 37067, USA\\
$^{17}$ CBA-France, 9 Rue Vasco de Gama, 59553 Lauwin Planque, France\\
$^{18}$ CBA-Orlando, Rolling Hills Observatory, 1643 Nightfall Drive, Clermont, 
FL, USA\\
$^{19}$ CBA-Mol, Andromeda Observatory, Oude Bleken 12, B-2400 Mol, Belgium\\
$^{20}$ CBA-Indiana, DeKalb Observatory H63, Auburn, IN 46706, USA\\
$^{21}$ College View Observatory, Warren Wilson College, Asheville, NC, USA\\
$^{22}$ CBA-Fresno, 1125 East Holland Avenue, Fresno, CA 93704, USA\\
$^{23}$ CBA-Newcastle, 9 Laking Drive, Newcastle, Ontario, Canada\\
$^{24}$ CBA-Finland, Hankasalmi Observatory, Verkkoniementie 30, FI-40950
Muurame, Finland\\
$^{25}$ 195 Rang 4 Ouest, St-Anaclet, QC, Canada G0K 1H0, Canada\\
$^{26}$ CBA-Concord, 1730 Helix Court, Concord, CA 94518, USA\\
$^{27}$ CBA-Cyprus, Green Island Observatory (B34), Gecitkale, North Cyprus\\
$^{28}$ Physics Department, Rochester Institute of Technology, Rochester, NY 14623,
USA\\
$^{29}$ Department of Physics, Middlebury College, Middlebury, VT 05753, USA\\
}

\pubyear{2015}

\begin{document}

\label{firstpage}
\pagerange{\pageref{firstpage}--\pageref{lastpage}} 
\maketitle

\begin{abstract}
We report the results of a long campaign of time-series photometry on the
nova-like variable UX Ursae Majoris during 2015. It spanned 150 nights, with
$\sim$ 1800 hours of coverage on 121 separate nights. The star was in its
normal `high state' near magnitude $V=13$, with slow waves in the light
curve and eclipses every 4.72 hours. 
Remarkably, the star also showed a nearly sinusoidal signal with a full amplitude
of 0.44 mag and a period of $3.680\pm 0.007$ d. We interpret this as the signature
of a retrograde precession (wobble) of the accretion disc. The same
period is manifest as a $\pm 33$ s wobble in the timings of mid-eclipse,
indicating that the disc's centre of light moves with this period.
The star also showed strong `negative superhumps' at frequencies
$\omega_{\rm orb}+N$ and
$2\omega_{\rm orb}+N$, where $\omega_{\rm orb}$ and $N$ are respectively
the orbital and precession frequencies. It is possible that these
powerful signals have been present, unsuspected, throughout the more
than 60 years of previous photometric studies.

\end{abstract}


\section{Introduction}
      UX Ursae Majoris (UX UMa) is one of the oldest and most thoroughly studied of 
the cataclysmic variables (CVs).  Among non-eruptive CVs, it's probably the 
champion in both respects.  Visual and photoelectric photometry showed it to be 
an eclipsing binary with a remarkably short period of 4.72 hours 
\citep{zverev37,johnson54,krzeminski63},
and \citet{walker54} proposed a model in which the hot star in the 
binary is surrounded by a large ring of gas on which a bright region 
(hot spot) resides.  The hot spot became a key feature of the basic model 
for understanding CVs, in which the spot is interpreted as the region where 
the mass-transfer stream impacts the outer edge of the accretion disc.

    The spectrum of UX UMa closely resembles that of dwarf novae in eruption: 
a blue continuum with broad, shallow hydrogen absorption lines, and narrow H 
emission contained within these absorption troughs.  
\ion{He}{i} and weak \ion{He}{ii} 
emission are sometimes also present.  
Recent spectroscopic studies have been 
reported by \citet{linnell08} and \citet{neustroev11}.  
The distance is $345\pm 34$ pc \citep{baptista95,baptista98}.
The out-of-eclipse mean $V$ magnitude is $\sim$13.0, but this is adversely 
affected by interstellar extinction ($\sim$0.2 mag) and the geometrical 
projection of a fairly edge-on disc 
($\sim$1.0 mag; \citealt{paczynski80}).
After these corrections, the angle-averaged 
$\langle M_V \rangle$ is about $+4.1$.  That's just about right for 
the `high state' of a dwarf nova with an orbital period of 4.7 hours 
(Fig.~1 of \citealt{joep11_distances}). 
Thus the spectrum and brightness are consistent 
with interpretation as a dwarf nova in the high state.

     In addition, UX UMa shows another phenomenon which is highly 
characteristic of dwarf novae: very rapid ($\sim$30 s) oscillations in its 
optical and UV brightness 
\citep{warner72,nather74,knigge98a}.
These oscillations are seen in practically every dwarf nova near the peak of 
eruption, and are consequently called `dwarf nova oscillations'
(DNOs; \citealt{joep81}, especially the abstract and Fig.~17).  
Their presence in UX UMa is yet another reason why the star is commonly 
regarded, and described, as essentially a `permanently erupting dwarf nova'. 

   UX UMa vaulted to the world's attention from a program of 
time-series photometry in the 1940s.  
We launched a more intensive program in 2015, and discovered several 
additional periodic signals, which we describe in this paper and interpret 
as signifying the retrograde precession of the accretion disc.

\section{Observations}

    We conducted this campaign with our global network of small photometric 
telescopes, the Center for Backyard Astrophysics (CBA).  The network's general 
approach to instrumentation and observing methods is given by 
\citet{skillman93}, and the summary observing log is given in Table~\ref{obs-log}.  
We used differential photometry with respect to one of the 
nearby field stars
GSC 3469-0356 ($V=13.068$),  
GSC 3469-0290 ($V=13.370$), and 
GSC 3469-0867 ($V=13.497$), with magnitudes corresponding to the
APASS photometric survey \citep{henden12}. 
We constructed light curves using overlaps of the various
time series to calibrate each on a common instrumental scale. 
That scale is roughly a $V$ magnitude
since most of our data is unfiltered in order to improve signal-to-noise.
Nevertheless, we did obtain sufficient data with a true $V$ filter to measure 
offsets, and this allowed us to place all our data into a magnitude scale
that is expected to nearly correspond to a true $V$, with a zero-point
uncertainty of $\sim$0.04 mag.

 The cycle time (integration + readout) between points in the various time
series was usually near $\sim$60 seconds. We made no correction for differential 
(color) extinction, although such a correction is in principle necessary, since 
all CVs are bluer than field stars. But in a long time series, such effects are 
always confined to the same frequencies (very near 1 and 2 cycles 
per sidereal day), so the resultant corruption is easily identified and 
ignored. In the present case, it is also mitigated by the northern latitudes 
of observers and the far-northern declination of the star (51 degrees), 
which made it possible to obtain long runs within our self-imposed limit of 
2.0 airmasses. Finally, we just prefer to keep human hands off the data as 
much as possible.
 
As detailed in Table~\ref{obs-log}, the campaign amounted to 355 separate time series 
on 121 nights
distributed over a span of 153 nights from February 24 to July 26, 2015. 
The total coverage was 1785 hours, 
all from sites in Europe and North America. 
This longitude span permitted many $\sim$14 hour runs, which eliminated all 
possibility of daily aliases 
-- the usual bugaboo of single-longitude time series.


\begin{table}
 \centering
\caption{
Log of observations.}
\centering
\begin{tabular}{llr}
\hline
\multicolumn{1}{c} {Observer}      &
\multicolumn{1}{c} {CBA station}   &
\multicolumn{1}{c} {Nights/hours} \\
\hline
Cejudo     &  Madrid (Spain) 0.3 m                 & 59/259 \\
Ulowetz    &  Illinois (USA) 0.24 m                & 53/216 \\
de Miguel  &  Huelva (Spain) 0.3 m                 & 31/173 \\
Jones      &  Oregon (USA) 0.35 m                  & 16/126 \\
Boardman   &  Wisconsin (USA) 0.3 m                & 19/104 \\
Barrett    &  Le Marouzeau (France) 0.2 m          & 24/99 \\
Koff       &  Colorado (USA) 0.25 m                & 14/95 \\
Stein      &  Las Cruces (New Mexico, USA) 0.35 m  & 12/85 \\
Campbell   &  Arkansas (USA) 0.15 m                & 15/76 \\
Vanmunster &  Belgium 0.35 m                       & 14/63 \\
Menzies    &  Massachusetts (USA) 0.35 m           &  9/60 \\
Slauson    &  Iowa (USA) 0.24 m                    & 15/59 \\
Goff       &  Sutter Creek (California, USA) 0.5 m &  9/48 \\
Roberts    &  Tennessee (USA) 0.4-0.5 m            &  8/45 \\
Morelle    &  France 0.3 m                         &  6/43 \\
Dvorak     &  Rolling Hills (Orlando, USA) 0.25 m  &  8/38 \\
Hambsch    &  Belgium 0.28 m                       &  7/29 \\
Starkey    &  Auburn (Indiana, USA) 0.4 m          &  4/28 \\
Collins    &  North Carolina (USA) 0.35 m          &  6/25 \\
Costello   &  Fresno (USA) 0.35 m                  &  4/21 \\   
M. Cook    &  Newcastle (Ontario, Canada) 0.4 m    &  4/20 \\
Oksanen    &  Finland 0.4 m                        &  3/20 \\
Lemay      &  Quebec (Canada) 0.35 m               &  7/24 \\
L. Cook    &  Concord (California, USA) 0.2-0.7 m  &  4/12 \\
Ogmen      &  Cyprus 0.35 m                        &  2/10 \\
Richmond   &  Rochester (New York, USA)  0.30 m    &  2/7 \\
\hline
\label{obs-log}
\end{tabular}
\end{table}

\section{Light curves and eclipses}

\begin{figure}
\includegraphics[angle=-90,width=\columnwidth]{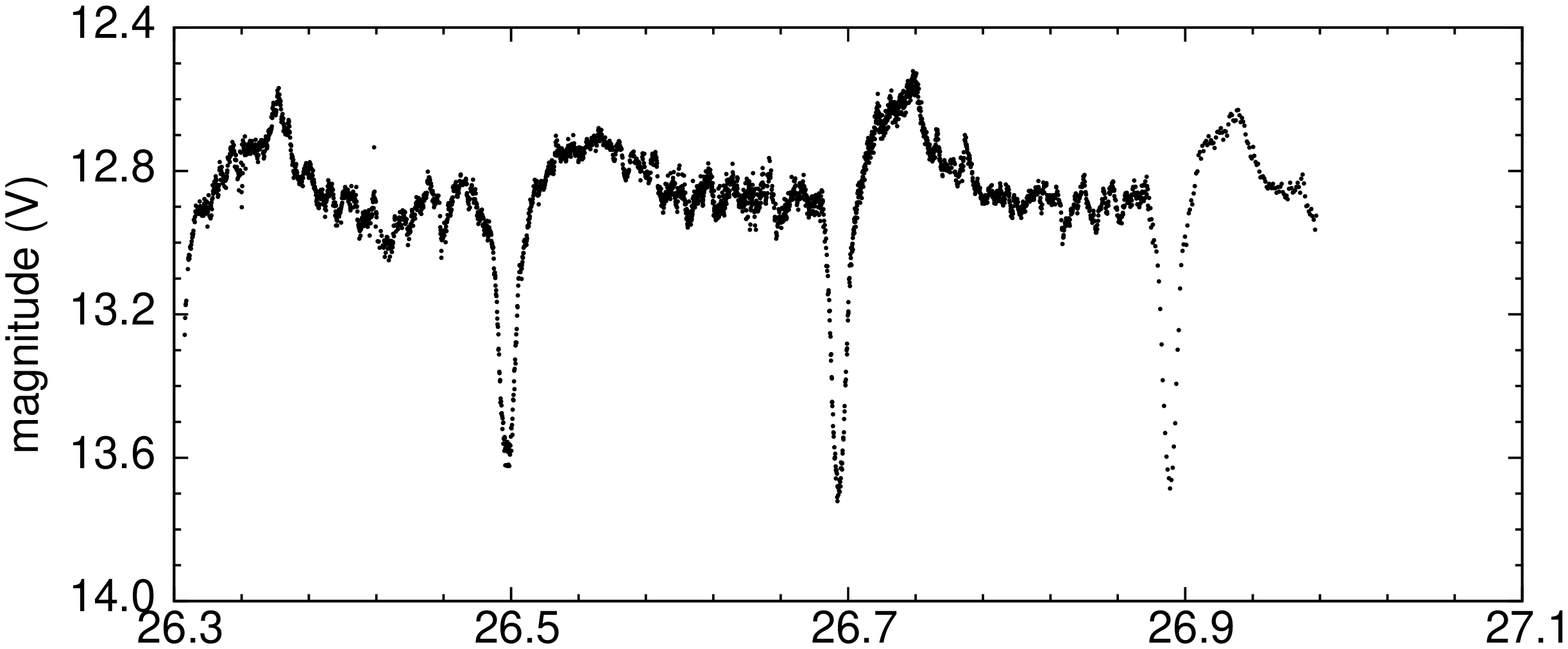}

\includegraphics[angle=-90,width=\columnwidth]{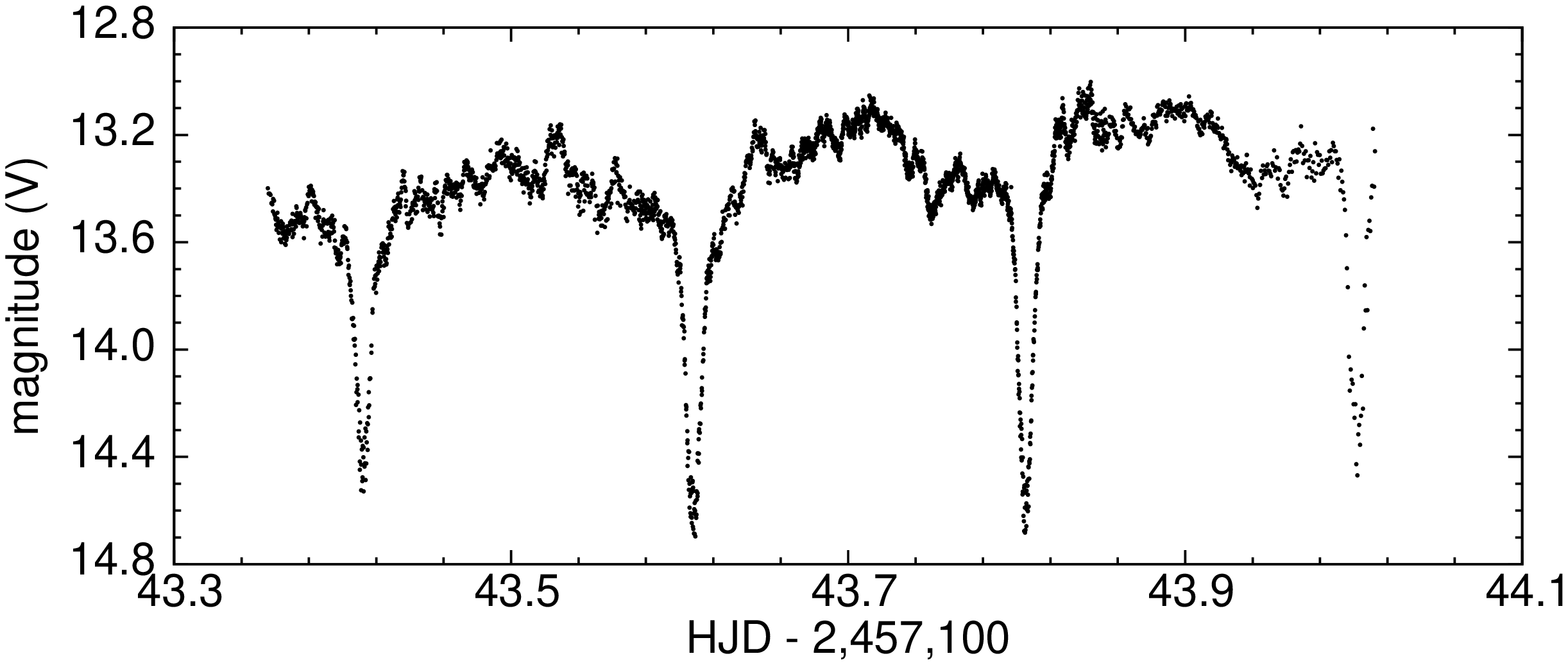}

\caption{
Representative light curves of UX UMa on two nights in the 2015 campaign.
}
\label{lcs-2nights}
\end{figure}

\begin{figure}
\includegraphics[angle=-90,width=\columnwidth]{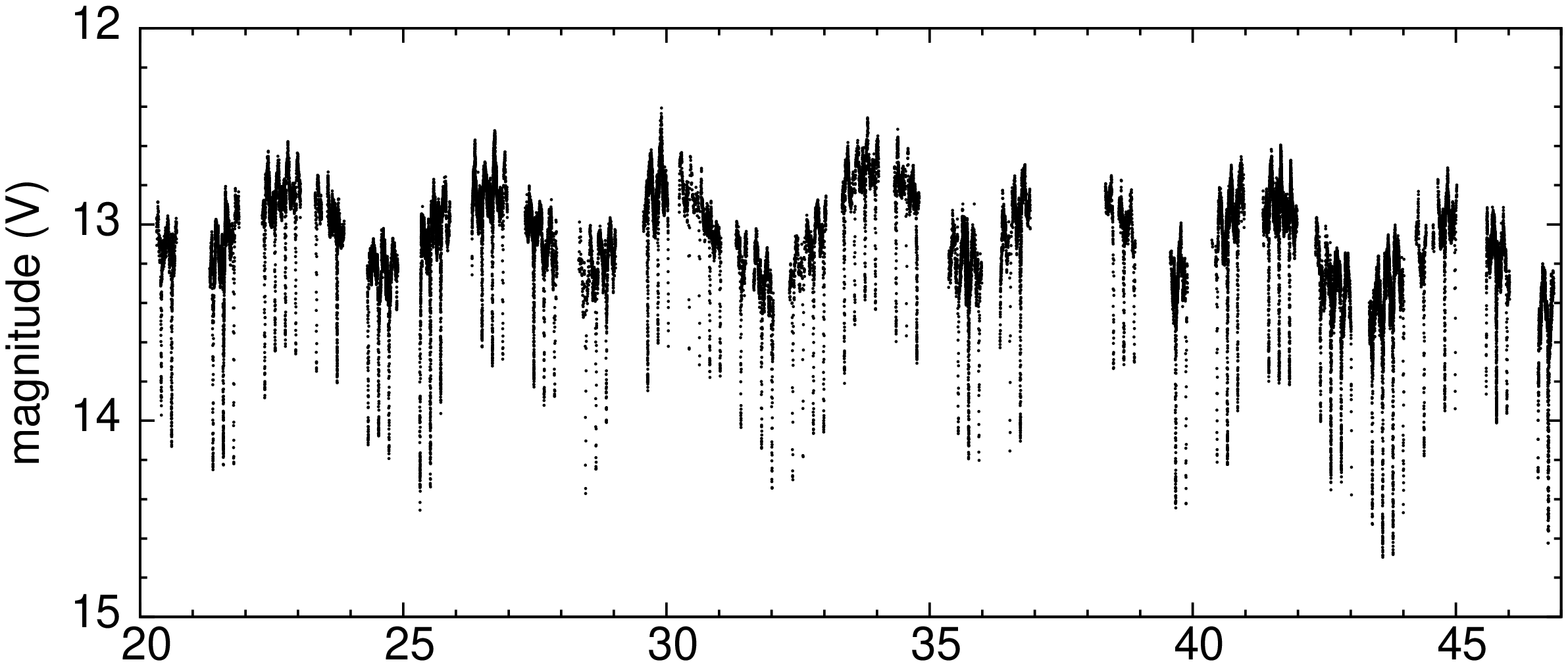}

\includegraphics[angle=-90,width=\columnwidth]{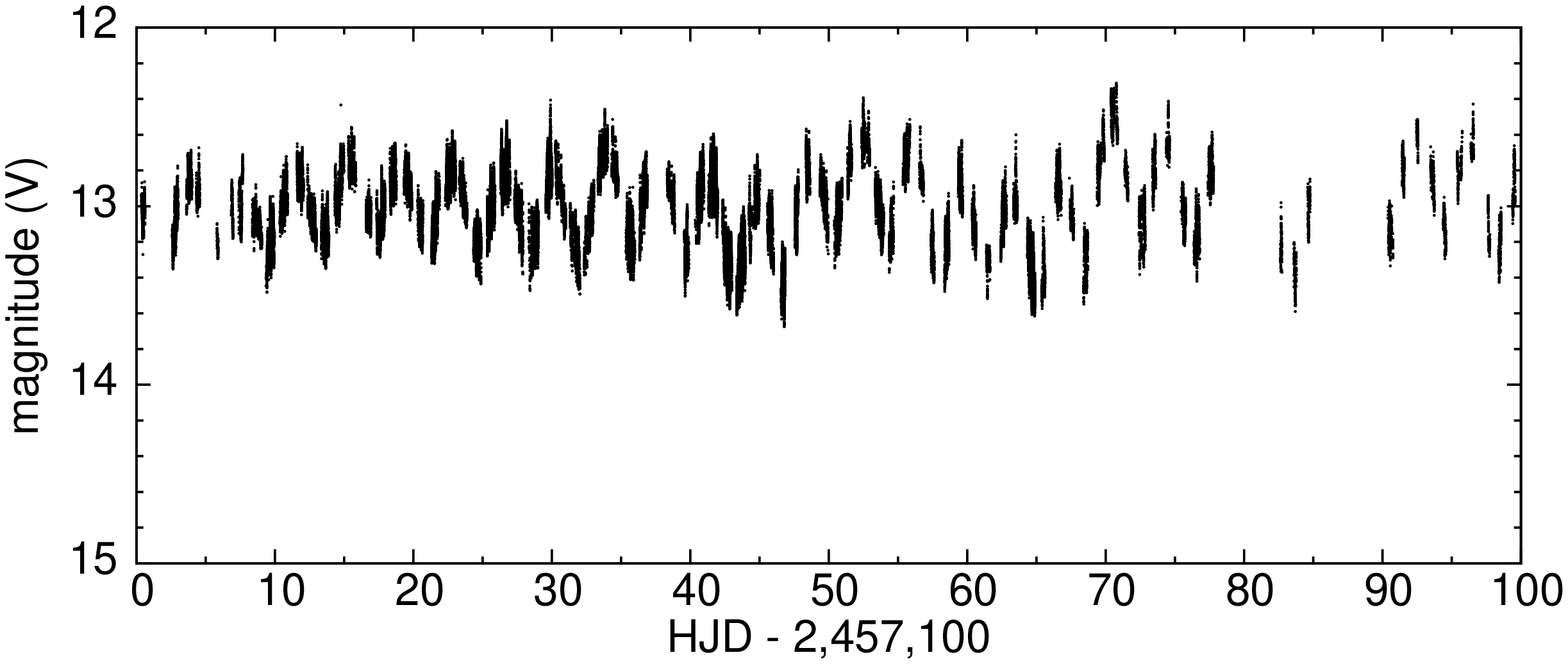}

\caption{
{\it Upper frame}: a 27-day light curve, showing eclipses, possibly
`orbital' humps, and a candidate $\sim$3.7 d variation (also
apparent in the eclipse minima). 
{\it Lower frame}: the central 100 days of the campaign, with eclipses
removed. The $\sim$3.7 d variation seems to endure throughout. 
}
\label{lc-ooe}
\end{figure}

Two representative nightly light curves are shown in Fig.~\ref{lcs-2nights}.
They are similar to essentially 
all light curves in the literature (e.g. \citealt{johnson54,walker54,warner72}): 
regular, asymmetric eclipses, with ingress being steeper than egress; 
irregular, non-coherent variations at short timescales 
(flickering); plus a roughly `orbital' hump, 
although the latter varies markedly -- and interestingly! -- from one night to 
the next. 
The mean brightness outside the eclipses and at minimum are 
$V=13.02$ and $V=13.94$, respectively. These values are far from constant,
and vary from one orbital cycle to the next. 
The upper frame of Fig.~\ref{lc-ooe} shows a sample 
27-day light curve, which 
suggests the presence of a slow wave with a period near 3.7 d that modulates
the out-of-eclipse brightness, as well as the magnitude of the system at
mid-eclipse.
And the bottom frame shows a 100-day light 
curve (with eclipses removed), which 
confirms the apparent stability of this slow wave.

\begin{figure}
\includegraphics[angle=-90,width=\columnwidth]{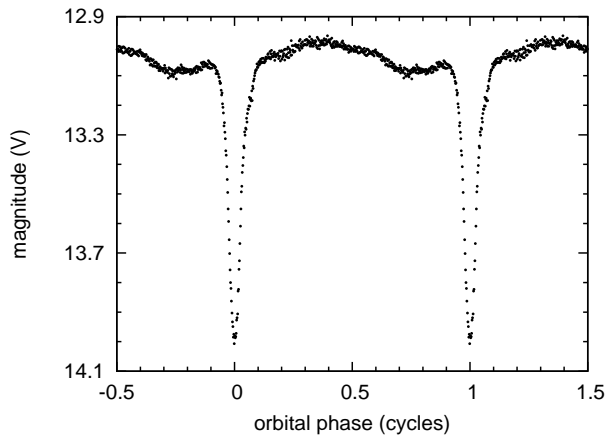}

\caption{
Mean orbital light curve over the full 5-month campaign. The mean
out-of-eclipse magnitude is $V=13.02$ mag and the mean eclipse
depth is 0.92 mag. Maximum light occurs near orbital phase 0.35.
}
\label{lc-orb}
\end{figure}

 We measured the time of minimum eclipse and the corresponding magnitude
by fitting a parabolic function to the bottom
half of the minimum ($\pm 0.04$ in orbital phase). Individual errors were
estimated by Monte Carlo methods and found to vary in the range 
$(0.7$--$6)\times 10^{-4}$ d, with median value of $2\times 10^{-4}$ d.
Strictly speaking, this fitting procedure provides 
estimates of the time of minimum light, which tends to occur slightly later
than the time of mid-eclipse in 
UX UMa \citep{baptista95}. But our data do not allow us to discriminate
between these two timings, since the differences are much smaller than 
our uncertainties. A total of 214 minima
were timed. These times, collected in Table~\ref{timings-mid}, were found to track the 
ephemeris

\begin{equation}
T_{\rm min} ({\rm HJD}) = 2,457,078.51002(6) + 0.19667118(19) \,E  \, .
\label{ephem_min}
\end{equation}

Not surprisingly, the corresponding $O-C$ residuals were found to show no
statistically significant departure from linearity over the $\sim$150 d
baseline, since the orbital modulation is expected to be a stable clock
on this time scale. But as we shall see below, they appear to be modulated by the
3.7 d period described above.

\begin{table*}
 \centering
\caption{Timings of mid-eclipse (${\rm HJD} - 2,457,000$).}
\centering
\begin{tabular}{rrrrrrrrrr}
\hline
  78.5100 &  78.7067 &  79.6903 &  81.8535 &  83.6238 &  83.8202 &  84.6072 &
  88.5405 &  88.7371 &  89.7204 \\  
  91.8847 &  93.6535 &  93.8504 &  95.8169 &  96.6035 &  97.5871 &  98.5704 &  
  99.3574 &  99.5543 &  99.7508 \\ 
 100.3406 & 100.5373 & 102.7010 & 102.8974 & 103.6839 & 103.8804 & 104.4702 & 
 105.8460 & 106.8306 & 107.4204 \\
 107.6177 & 108.4037 & 108.6006 & 108.7971 & 108.9946 & 109.3872 & 109.5843 &
 109.7808 & 109.7807 & 109.9780 \\
 110.3708 & 110.5677 & 110.7644 & 111.7466 & 111.9439 & 112.3373 & 112.5341 & 
 112.7302 & 112.9277 & 113.5175 \\
 113.7146 & 114.5015 & 114.6975 & 114.8945 & 115.4837 & 116.6632 & 116.8615 & 
 117.6474 & 117.8443 & 118.4344 \\
 118.6303 & 119.4173 & 119.6142 & 119.8106 & 120.4012 & 120.5982 & 121.3848 & 
 121.5817 & 121.7782 & 122.3679 \\
 122.5644 & 122.7605 & 122.9570 & 123.3509 & 123.7435 & 124.3346 & 124.5313 &
 124.7283 & 125.3184 & 125.5147 \\
 125.7115 & 126.4977 & 126.6944 & 126.8909 & 127.4810 & 127.6765 & 128.4647 & 
 128.6618 & 128.8575 & 129.6447 \\
 129.8404 & 130.4308 & 130.6275 & 130.8235 & 131.0210 & 131.4147 & 131.8079 & 
 132.0053 & 132.3981 & 132.5946 \\
 132.7920 & 132.9887 & 133.3812 & 133.5772 & 133.7745 & 133.9711 & 134.3641 & 
 134.5612 & 134.7576 & 135.5452 \\
 135.7413 & 135.9382 & 136.5286 & 136.7253 & 138.4939 & 138.6909 & 138.8884 & 
 139.6751 & 139.8719 & 140.4620 \\
 140.6585 & 140.8542 & 141.4447 & 141.6413 & 141.8378 & 142.4276 & 142.6249 & 
 142.8214 & 143.4122 & 143.6082 \\
 143.8053 & 144.0018 & 144.3949 & 144.7883 & 145.7714 & 145.9683 & 146.7553 & 
 147.7391 & 148.5249 & 149.5079 \\
 149.7052 & 150.4920 & 150.6885 & 150.8851 & 151.4758 & 151.6720 & 152.4582 &
 152.8514 & 153.4418 & 153.6377 \\
 153.8349 & 154.4253 & 154.6217 & 155.4089 & 155.6054 & 155.8012 & 156.5882 & 
 156.7845 & 157.5718 & 158.5556 \\
 159.5390 & 160.5210 & 162.4896 & 163.4720 & 164.4554 & 164.8488 & 165.4389 & 
 166.4216 & 166.6187 & 166.8156 \\
 167.4055 & 167.6011 & 168.3887 & 168.5853 & 169.5697 & 169.7657 & 170.5522 & 
 170.7485 & 171.5346 & 172.5197 \\
 173.5027 & 175.4691 & 175.6662 & 176.6494 & 177.4359 & 177.6320 & 183.7289 & 
 184.7127 & 190.4157 & 190.6130 \\
 191.4001 & 192.5786 & 193.5618 & 194.5464 & 195.7266 & 196.5122 & 197.6937 & 
 198.4796 & 199.4623 & 201.6262 \\
 201.8226 & 202.4132 & 203.7890 & 206.5436 & 208.5088 & 209.4941 & 209.6898 & 
 211.6558 & 221.4904 & 222.4728 \\
 223.4569 & 224.4399 & 227.3903 & 227.7846 &          &          &          &
          &          &         \\
\hline
\end{tabular}
\label{timings-mid}
\end{table*}

 The orbital light curve is significantly contaminated by flickering, the 3.7 d
modulation, and the `superhump' variations described below. Making no attempt 
to remove these effects, and simply averaging over the $\sim$ 1800 hours of 
coverage, 
we found the mean orbital light curve seen in Fig.~\ref{lc-orb}. This appears to be the 
first mean orbital light curve published for this venerable, oft-observed star. 
And it shows maximum light near orbital phase 0.35, 
roughly 180$^\circ$ out of phase with the standard `hot spot' model developed for 
U Geminorum, and thought to prevail, {\it mutatis mutandis}, in all CVs 
\citep{smak71,warner71}. The accretion geometry must be significantly 
different in UX UMa.

\begin{figure}
\includegraphics[angle=-90,width=\columnwidth]{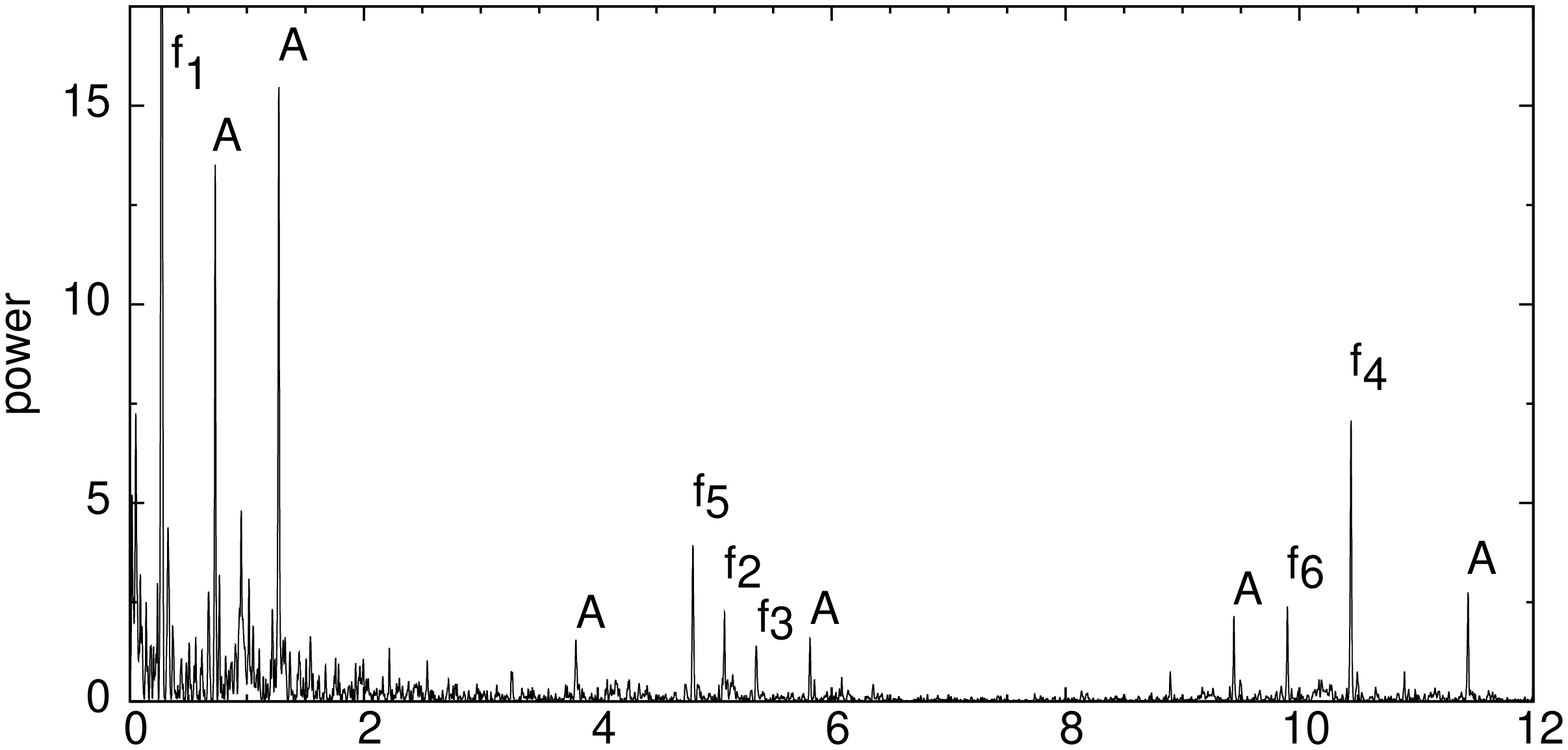}

\vspace*{-1.75cm}
\includegraphics[angle=-90,width=\columnwidth]{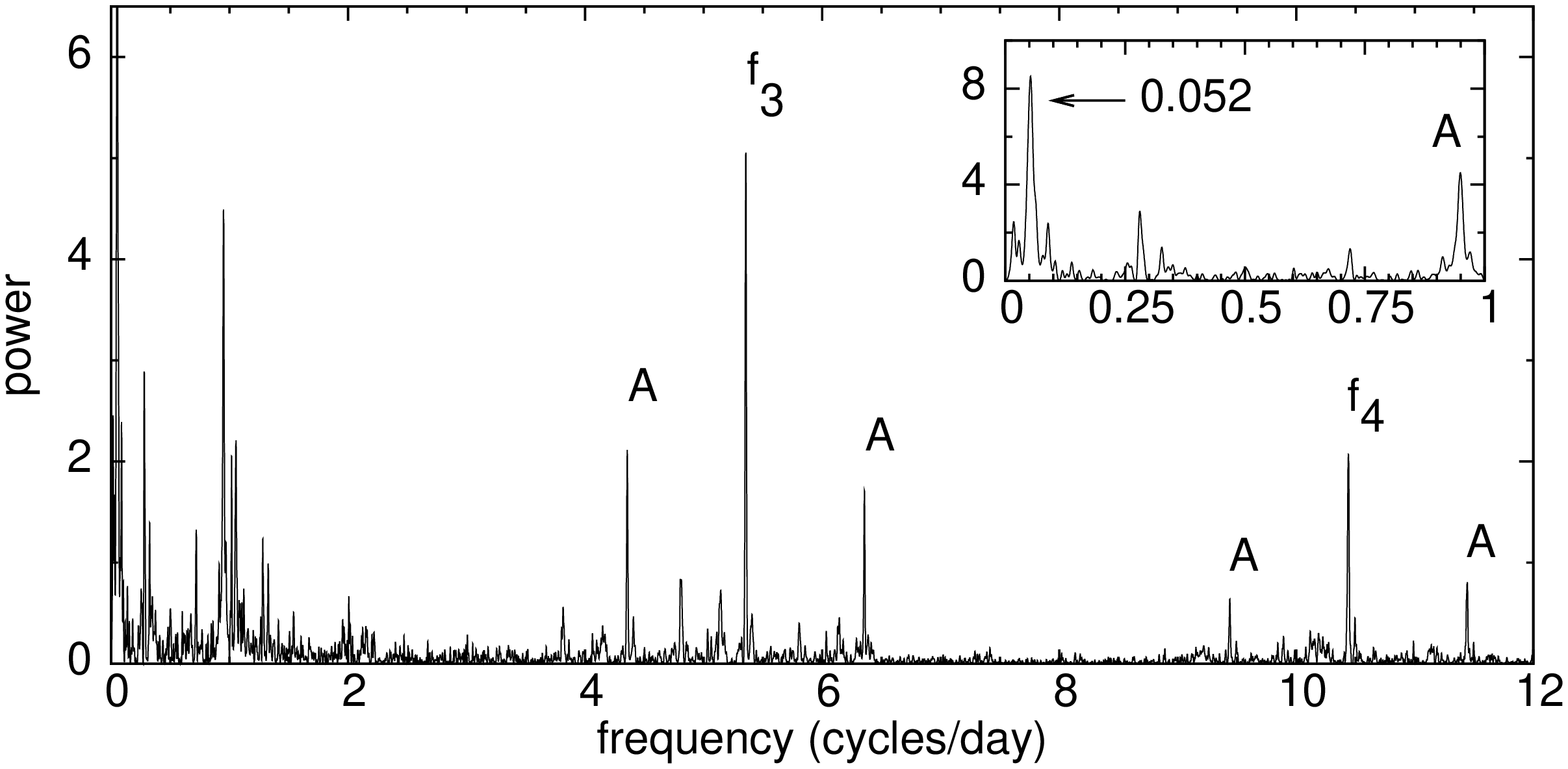}

\caption{
{\it Upper frame}:
power spectrum of the full, 5-month light curve (with eclipses removed).
The most prominent peaks are labeled $f_1$ through $f_6$, and one-day
aliases are designated `A'. There are only two independent (unrelated)
frequencies: the strong signal 
$f_1=0.2717$ c d$^{-1}$ (nodal, $N$), which 
rises off-scale to a 
power of 43.5
(or semi-amplitude of 0.22 mag); and $f_2$, which
corresponds to the orbital frequency, $\omega_{\rm orb}$.
{\it Lower frame}:
power spectrum of the residual light curve, after the nodal and orbital
frequencies are subtracted. The two obvious peaks ($f_3$ and $f_4$) occur
at $\omega_{\rm orb} + N$ and $2\omega_{\rm orb} + N$, and correspond
to `negative superhumps'. The strong peaks labeled $f_5$ and $f_6$
in the upper frame coincide with $\omega_{\rm orb} - N$ and $2\omega_{\rm orb} - N$.
These probably arise from modulation of the orbital signal by $N$, and 
their amplitude is greatly reduced
after the subtraction. Inset is a zoom-in of the power spectrum
in the range 0--1 c d$^{-1}$ showing a possible detection at 0.0521(8) c d$^{-1}$
(see text for details).
}
\label{p04-ooe}
\end{figure}

\section{Periodic signals in the light curve}

 Our primary analysis tool for studying periodic waves is power spectra
calculated by Fourier methods. 
The frequency analysis was performed by using the {\sc Period04} package
\citep{period04}, based on the discrete Fourier transform method.
Uncertainties in the frequencies and amplitudes were estimated by using
Monte Carlo methods from the same package. Of course the sharp eclipses 
severely contaminate analysis by Fourier methods, since the latter represent 
time series as sums of sinusoids. So to prepare the light curves for study, we 
first removed the eclipse portion of the light curves, viz. the phase interval 
0.9--1.1. 

\begin{figure}
\includegraphics[angle=-90,width=\columnwidth]{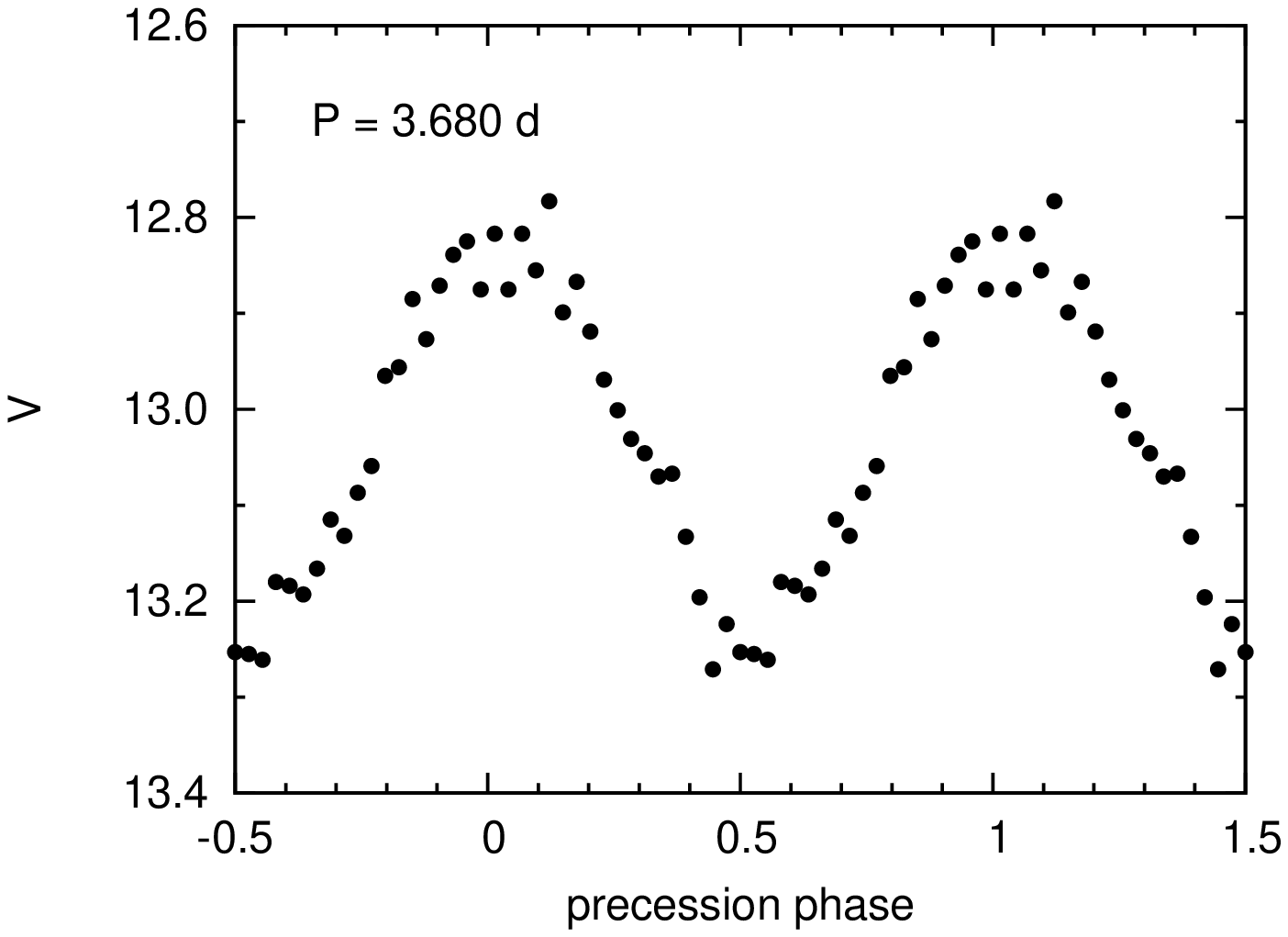}

\includegraphics[angle=-90,width=\columnwidth]{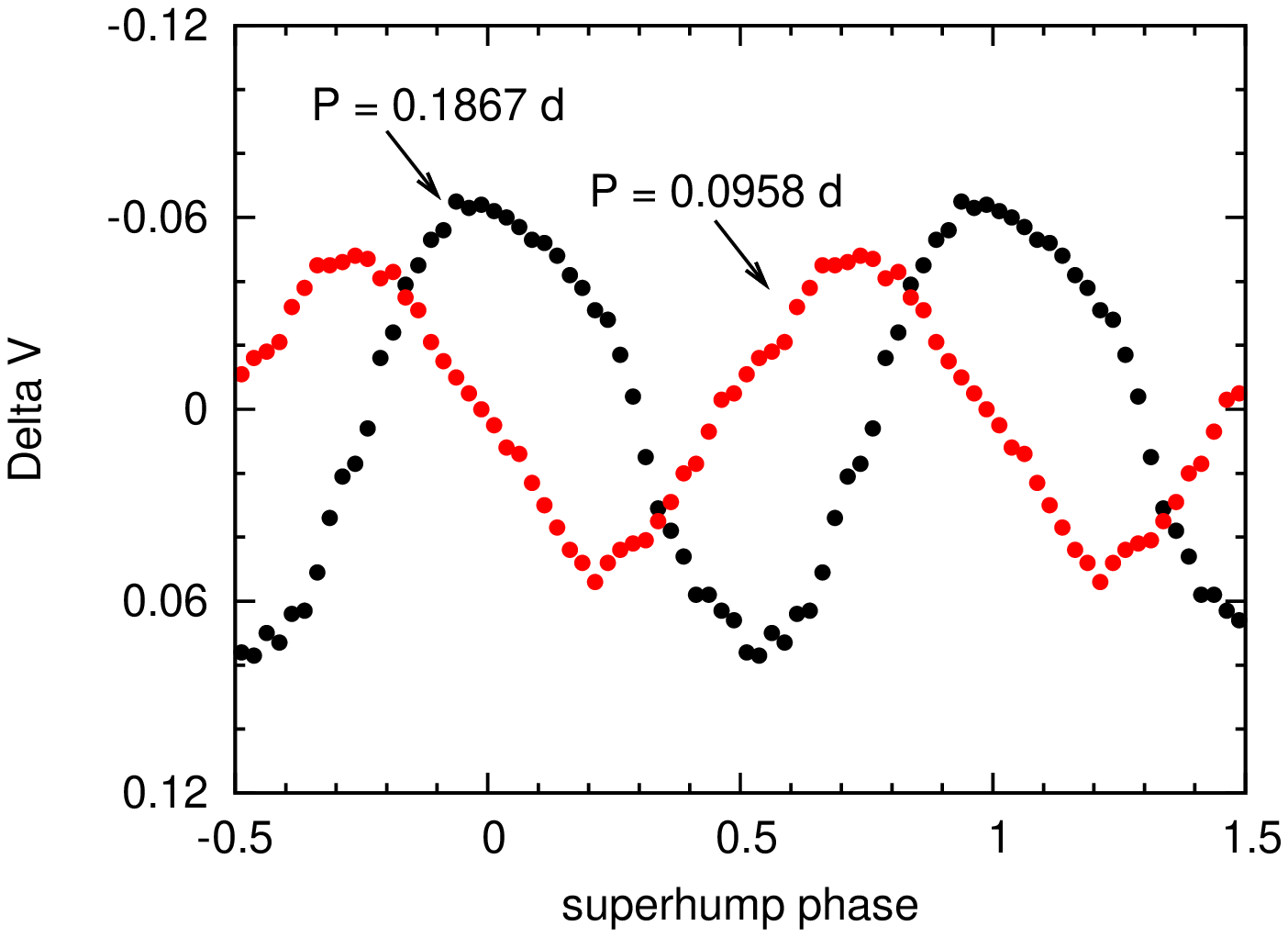}

\caption{
{\it Upper frame}: mean light curve folded on the 
nodal ($N$) frequency (0.2717 c d$^{-1}$ or 3.68 d) relative
to the ephemeris given in Eq.~(\ref{ephem_nodal}).
{\it Lower frame}: mean light curve after removing the nodal and orbital
signals folded on the $\omega_{\rm orb} + N$ frequency 
(5.3562 c d$^{-1}$ or 0.1867 d, black points) and the 
$2\omega_{\rm orb} + N$ frequency (10.4391 c d$^{-1}$ or 0.0958 d,
red points) relative to the ephemeris given in Eq.~(\ref{ephem_nsh}).
}
\label{waveforms}
\end{figure}

The low-frequency region of the power spectrum is shown in the upper
frame of Fig.~\ref{p04-ooe}, 
where the most prominent peaks are 
labeled $f_1$ through $f_6$, 
and alias peaks marked with `$A$'. 
In the figure, and throughout the paper, frequencies are expressed
in cycles per day,\footnote{The 
natural frequency unit for time-series studies on a planet plagued by rotation
and sunrise.} for which we use c d$^{-1}$ as a shorthand.
A prominent peak ($f_2$ in Fig.~\ref{p04-ooe}) is observed at 
5.0847(9) c d$^{-1}$. This signal, with a semi-amplitude of 0.042 mag,
coincides with the orbital frequency $\omega_{\rm orb}$.
But the most powerful signal ($f_1$ in Fig.~\ref{p04-ooe}) 
occurs at 0.2717(5) c d$^{-1}$, or 3.680(3) d, a signal
that is unrelated to the orbital motion, and that we denote
as $N$, in anticipation of identifying it with nodal 
precession of the accretion disc. We summed at 0.2717 c d$^{-1}$, and found
a highly sinusoidal waveform with a semi-amplitude of 0.22 mag. This is shown
in the upper frame of Fig.~\ref{waveforms}.

In addition to $\omega_{\rm orb}$ and $N$, other signals appear in the vicinity of 
$\omega_{\rm orb}$ and $2\omega_{\rm orb}$. For their characterization, we 
subtracted the sinusoids corresponding to $N$ and $\omega_{\rm orb}$ from 
the full out-of-eclipse data set, and then recalculated the power spectrum
of the residual light curve. 
The results are shown in the lower frame of Fig.~\ref{p04-ooe}, which 
reveals obvious signals at 5.3562(4) and 10.4391(4) c d$^{-1}$ 
($f_3$ and $f_4$ in Fig.~\ref{p04-ooe}), with semi-amplitudes
of 0.069 and 0.034 mag, respectively. These are 
consistent with identifications as 
$\omega_{\rm orb}+N$ and $2\omega_{\rm orb} + N$, which are expected at 
5.3564(7) and 10.4410(9) c d$^{-1}$, respectively. 
These upper sidebands of the orbital frequency are
known as {\it negative superhumps} in variable-star nomenclature, because
in period (rather than frequency) language, their period excesses over
$P_{\rm orb}$, $P_{\rm orb}/2$, etc. are {\it negative}.\footnote{The 
terminology goes back to \citet{joep95-v503cyg}, and the full suite of CV 
periodic-signal arcana is reviewed in Appendix A of 
\citet{joep02_wzsge}.}
The mean waveform of these negative superhumps are shown in the lower
frame of Fig.~\ref{waveforms}.

The transition from the upper to the lower frame in Fig.~\ref{p04-ooe} looks odd. 
Of course the one-day aliases, 
along with the main peaks, disappear when the $N$ and $\omega_{\rm orb}$ 
signals are subtracted from the time series.  But in Fig.~\ref{p04-ooe} 
there are also strong peaks at 4.814(1) and 9.899(1) c d$^{-1}$ 
($f_5$ and $f_6$ in Fig.~\ref{p04-ooe}), with amplitudes 
greatly reduced after the subtraction.
That's surprising.  But these frequencies are essentially equal to
$\omega_{\rm orb}-N$ and $2\omega_{\rm orb}-N$, so a good possibility is 
that the dominant $N$ signal severely modulates the orbital signal, 
producing artificial flanking peaks at $\pm N$.  The effects 
described below in \S 5 support this.  Only the higher-frequency
$+N$ sidebands -- the negative superhumps -- survive the subtraction.
A summary of the main frequencies is given in Table~\ref{freqs}. 
 
 The power spectrum in the lower frame of 
Fig.~\ref{p04-ooe} seems to show a strong broad peak around 1 c d$^{-1}$.
This peak, centred around 0.948(2) c d$^{-1}$, is actually an alias of
a stronger detection at 0.0521(8) c d$^{-1}$, as shown inset in a
zoomed-in view of the power spectrum in the range 0--1 c d$^{-1}$. Is this 
detection an indication of a $\sim 19$ day periodicity in UX UMa? It could be,
but after a closer inspection we find no trace of this signal during, approximately,
the first half of the campaign. Admittedly, we have no grounds for believing that
this is a true detection, and we are more inclined to guess that it is just noise.

\begin{table}
 \centering
\caption{
The most significant frequencies, along with their semi-amplitudes
$(A)$ and physical interpretation.
}
\centering
\begin{tabular}{llll}
\hline
\multicolumn{1}{c} {label}      &
\multicolumn{1}{c} {frequency}  &
\multicolumn{1}{c} {$A$}  &
\multicolumn{1}{c} {meaning}    \\
     & 
\multicolumn{1}{c} {(c d$^{-1}$)}  &  
\multicolumn{1}{c} {(mag)}         &   \\
\hline
$f_1$  &  0.2717(5) & 0.220(1) &  $N$ (nodal) \\
$f_2$  &  5.0847(9) & 0.042(1) &  $\omega_{\rm orb}$ (orbital) \\
$f_3$  &  5.3562(4) & 0.069(1) &  $\omega_{\rm orb}+N$ (nsh) \\
$f_4$  & 10.4391(4) & 0.034(1) &  $2\omega_{\rm orb}+N$ (nsh) \\
$f_5$  & 4.814(4)   &          &  $\omega_{\rm orb}-N$       \\
$f_6$  & 9.899(4)   &          &  $2\omega_{\rm orb}-N$       \\
\hline
\label{freqs}
\end{tabular}
\end{table}

The waveforms of all four physically significant signals 
($N$, $\omega_{\rm orb}$,
$\omega_{\rm orb}+N$, $2\omega_{\rm orb}+N$)
are impressively sinusoidal, 
and probably indicate that none of these signals rely on the deep eclipse for 
their existence.  UX UMa would probably show these effects at any binary 
inclination, although the amplitude may well depend on inclination.

\begin{figure}
\includegraphics[angle=-90,width=\columnwidth]{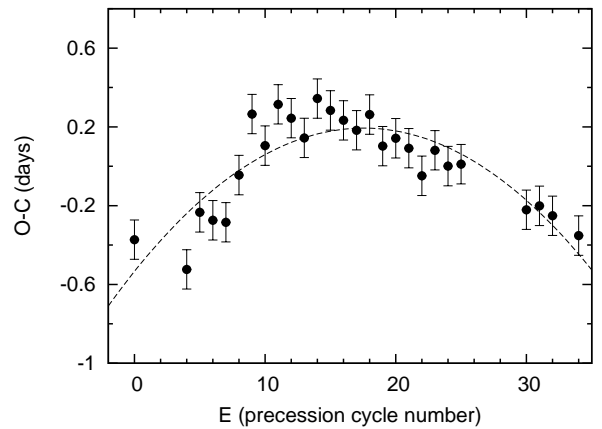}

\caption{
$O-C$ residuals of the timings of maximum light (Table~\ref{timings_nodal}) on the 
3.68 day cycle as determined from 
the ephemeris given in Eq.~(\ref{ephem_nodal}). The dashed curve represents the best
quadratic fit, from which we infer that the nodal period changes
by $\sim 0.2$ d over the 5-month campaign.}
\label{oc-nodal}
\end{figure}

\begin{table}
 \centering
\caption{Times of maximum light on the
3.68 day cyle (${\rm HJD} - 2,457,000$).}
\centering
\begin{tabular}{rrrrrrr}
\hline
  81.64 &  96.25 & 100.23 & 103.88 & 107.56 & 111.49 & 115.49 \\
 119.02 & 122.92 & 126.54 & 130.13 & 134.02 & 137.65 & 141.29 \\
 144.93 & 148.70 & 152.23 & 155.96 & 159.60 & 163.15 & 166.97 \\
 170.58 & 174.28 & 192.50 & 196.21 & 199.85 & 207.13 &         \\
\hline
\label{timings_nodal}
\end{tabular}
\end{table}

\subsection{The 3.7 d clock}

We have estimated the timings of maximum light in the 3.7 d cycle.
A total of 27 maxima were timed, with 
estimated uncertainties on individual timings of $\sim$0.1 d. 
These values are presented in Table~\ref{timings_nodal}. 
A linear regression to these timings provides the following test ephemeris

\begin{equation}
T_{\rm{max}} ({\rm HJD}) = 2,457,082.01(10) + 3.690(10) \,{\rm E} \, 
\label{ephem_nodal}
\end{equation}

\noindent
which we used to calculate the $O-C$ diagram shown in Fig.~\ref{oc-nodal}.
The curvature indicates that no constant period satisfies the data, but
rather a period drifting about a mean value of 3.69 d. From a 
parabolic fit to the $O-C$ residuals, we find 
that the period drifts at a rate of $dP/dt=-0.0013(2)$, or
$dN/dt = 9.5(1.5) \times 10^{-5}$ c d$^{-2}$, which amounts
to an overall decrease in the nodal period 
of $\sim$0.2 d over the 5 months spanned by our observations.

\begin{figure}
\includegraphics[angle=-90,width=\columnwidth]{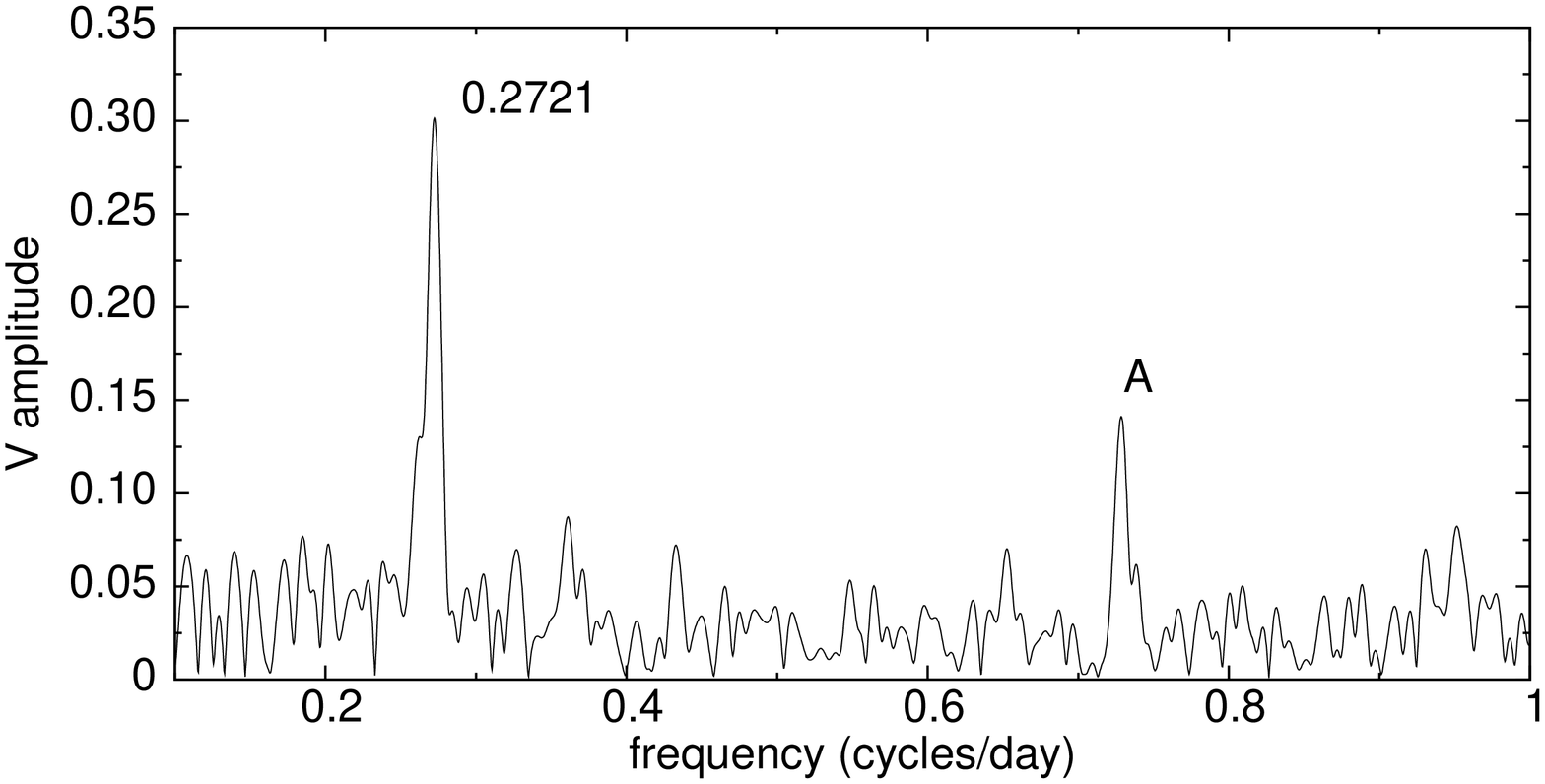}

\includegraphics[angle=-90,width=\columnwidth]{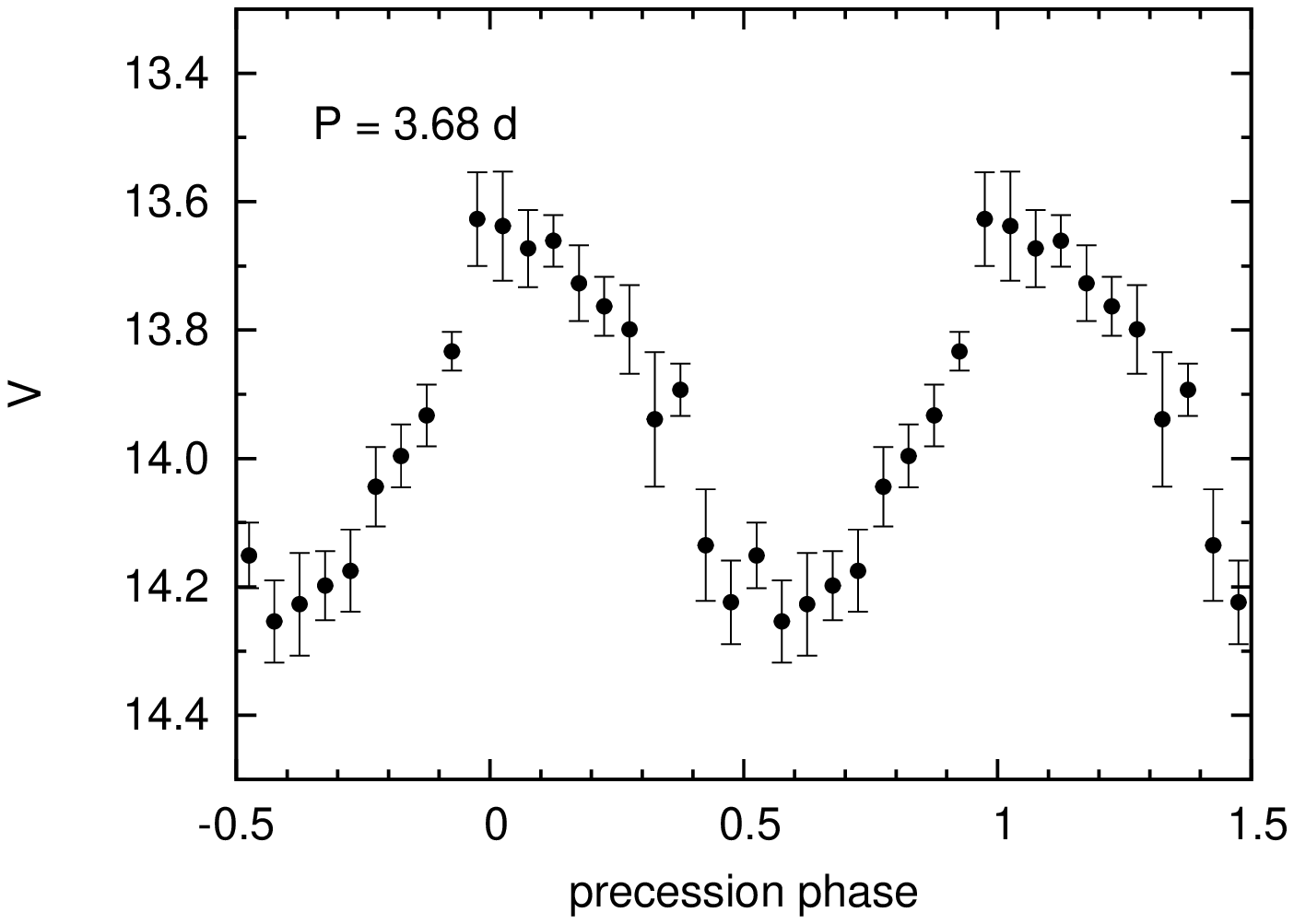}

\caption{
{\it Upper frame}:
power spectrum of the values of brightness at minimum light. The strongest
signal occurs at 0.2721(10) c d$^{-1}$, the nodal ($N$) frequency. 
The peak labelled with `A' is an alias centred at frequency $1-N$.
{\it Lower frame}: the values of brightness at minimum light folded on
the nodal frequency according to the ephemeris given in Eq.~(\ref{ephem_nodal}).}
\label{p04-min}
\end{figure}

\subsection{Periodic effect in the eclipse depths}

 As one may notice in Figs.~\ref{lcs-2nights} and \ref{lc-ooe}, 
the brightness at minimum light
($V_{\rm min}$) is notoriously variable, and presumably modulated by
the 3.68 d wave discussed earlier for the out-of-eclipse brightness
($V_{\rm out}$). 
As for the eclipse depths, variations
-- if they exist at all -- are not easy to perceive from these figures.
We tackle these questions next.

 We started by considering our
estimates of $V_{\rm min}$ 
and checked for possible periodic variations, finding the power spectrum 
shown in the upper frame of Fig.~\ref{p04-min}. 
The dominant peak at 0.2721 c d$^{-1}$ shows that 
the 3.68 d period also modulates the minimum light. The lower frame of 
Fig.~\ref{p04-min}
shows these values folded (and binned)
on the ephemeris given in Eq.~(\ref{ephem_nodal}), and indicates that $V_{\rm min}$
varies essentially as a sinusoid. Comparison with the out-of-eclipse
modulation $V_{\rm out}$ indicates that:
(i) both share the same periodicity;
(ii) $V_{\rm min}$ has a larger semi-amplitude (0.30 mag) than
$V_{\rm out}$ (0.22 mag); and
(iii) both are in phase (difference in phase of $0.02 \pm 0.03$, according
the corresponding best sinusoidal fits).

 What about the {\it eclipse depth}? If this is defined as
$\Delta V = V_{\rm min} - V_{\rm out}$, our previous analysis shows
that $\Delta V$ is indeed non-constant, and varies sinusoidally throughout 
the precession cycle. Adopting our best sinusoidal fits for $V_{\rm min}$
and $V_{\rm out}$, the eclipse depth is found to vary, on average, from 
0.82 mag at the maximum of the 
precession cycle (phase $\phi_p = 0$) to 1.00 mag at its 
minimum ($\phi_p = 0.5$). 

\begin{figure}
\includegraphics[angle=-90,width=\columnwidth]{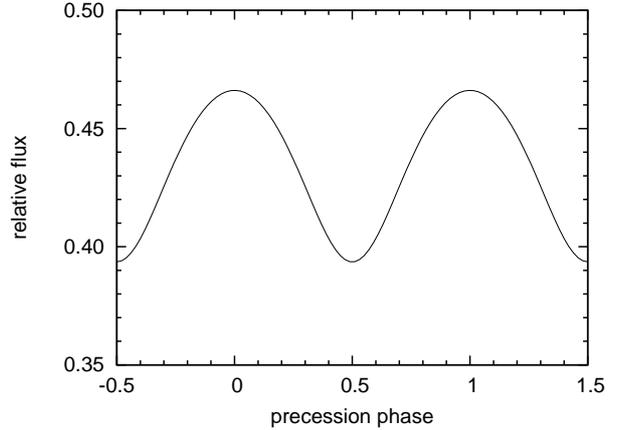}

\caption{
Relative flux, defined as the ratio between the flux at minimum light over
the out-of-eclipse flux, as determined from best sinusoidal fits, showing that
orbital eclipses are deepest at the minimum of the precession cycle.}
\label{flux-rat}
\end{figure}

 The cyclic effect on the eclipse depth can also be analyzed in terms
of fluxes. Fig.~\ref{flux-rat} shows the 3.68 day modulation of the flux at minimum light 
relative to the out-of-eclipse flux. In line with the discussion above in terms
of magnitudes, eclipses are deepest at the minimum of the precession cycle.
We note that a similar effect has been reported for the SW Sex nova-like PX And
\citep{stanishev02}.

\subsection{Periodic effect in the mid-eclipse residuals}

\begin{figure}
\includegraphics[angle=-90,width=\columnwidth]{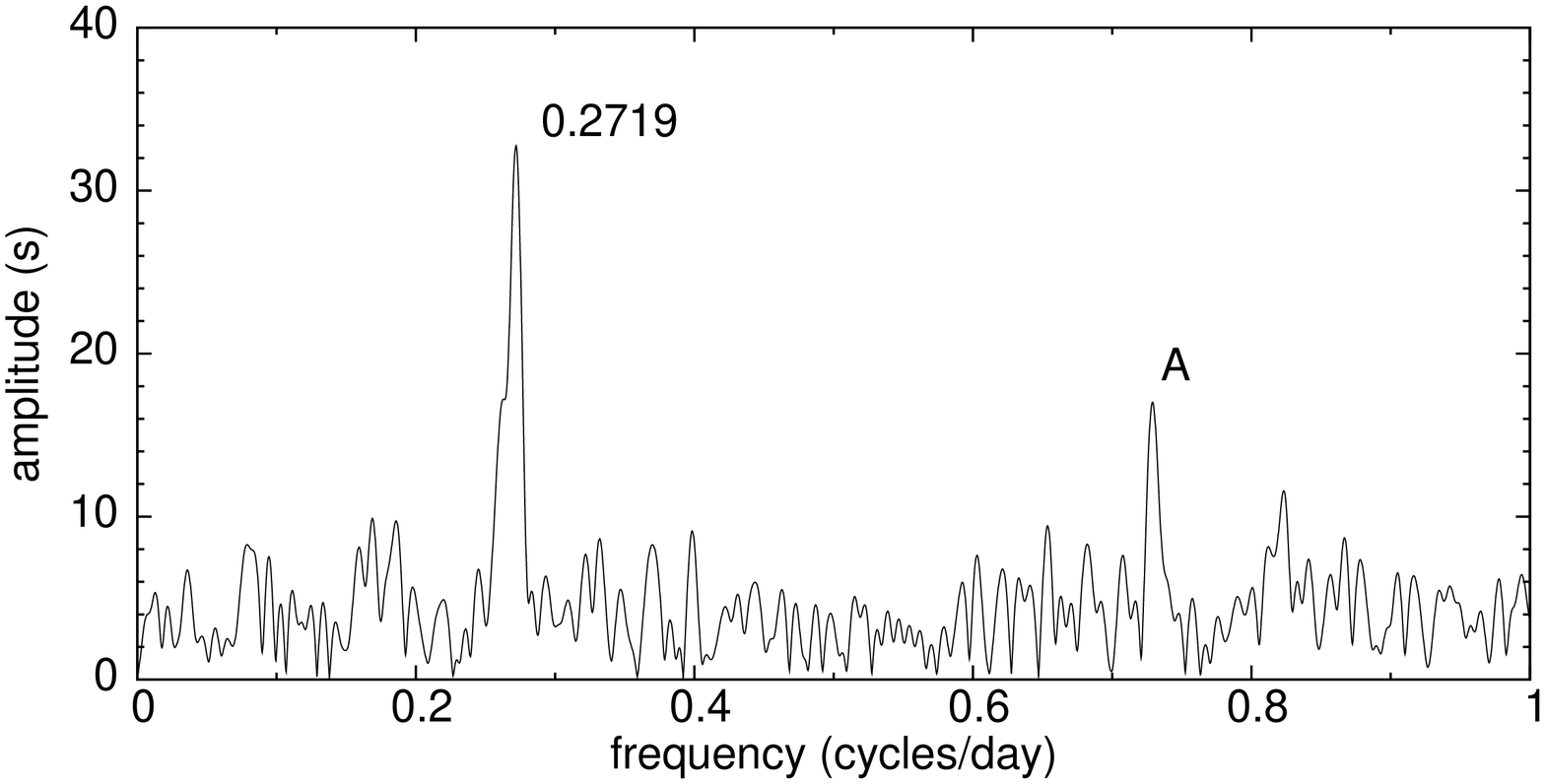}

\includegraphics[angle=-90,width=\columnwidth]{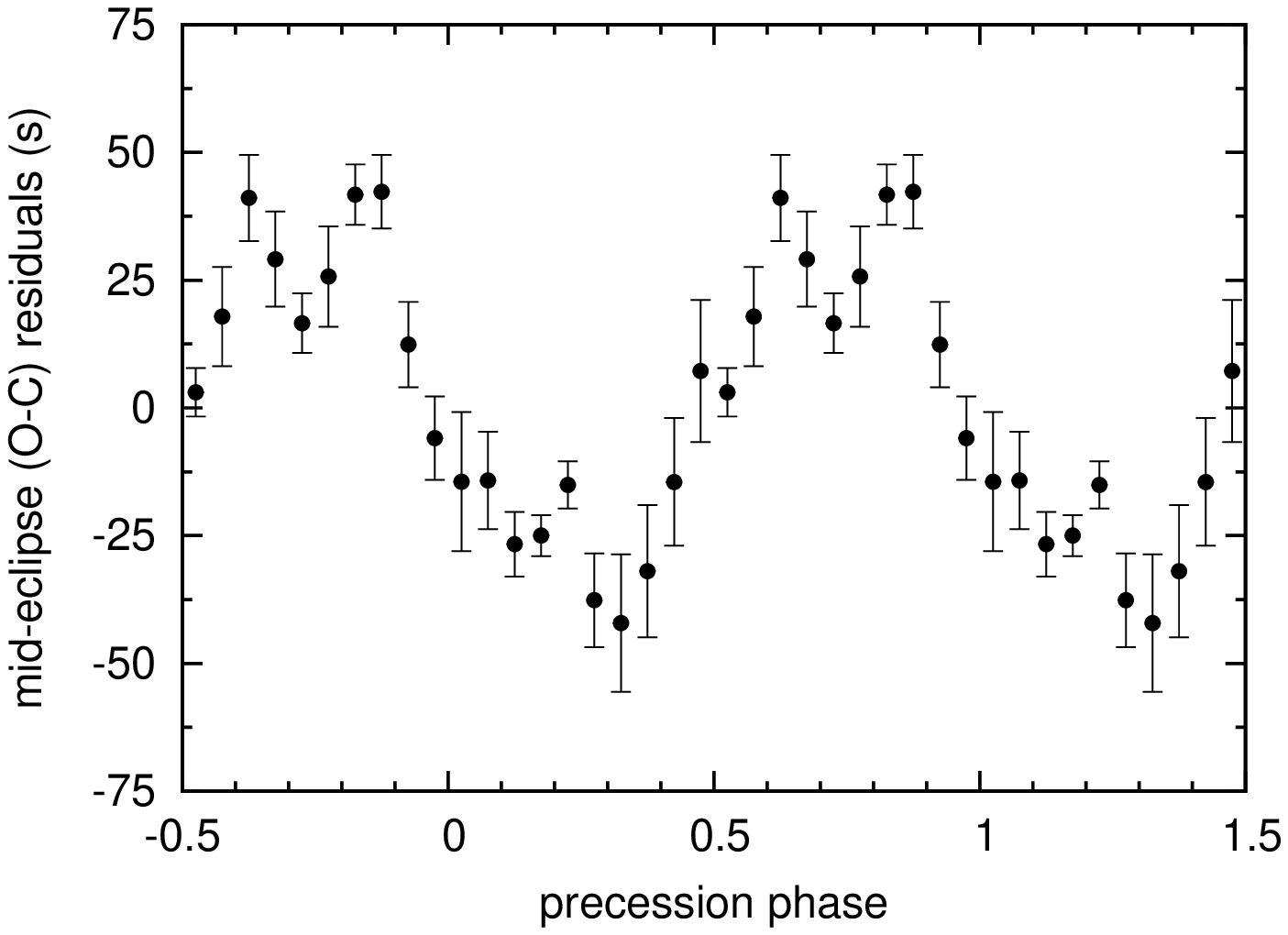}

\caption{
{\it Upper frame}:
power spectrum of the departures of eclipse timings from the ephemeris
given in Eq.~(\ref{ephem_min}). A significant peak
occurs at 0.2719(10) c d$^{-1}$, the same (nodal) frequency, $N$, 
characterizing the large
variations in light seen in Fig.~\ref{lc-ooe}. The peak labelled
with 'A' is an alias centred at frequency $1-N$.
{\it Lower frame}: fold of these residuals on 
the nodal frequency according to the ephemeris given in Eq.~(\ref{ephem_nodal}),
showing a periodic effect with a semi-amplitude of 33.3(9) s.}
\label{p04-res}
\end{figure}

As we examined the many eclipses, we noticed some which were distinctly
asymmetric, confounding the effort to derive a precise timing of mid-eclipse.
Departures from the mean ranged up to $\sim$80 s, but seemed to be systematic
with time. So we calculated the power spectrum of the departures of eclipse
timings from 
the ephemeris given in Eq.~(\ref{ephem_min}), and found the result seen in the upper 
frame of Fig.~\ref{p04-res}. 

A significant peak is present at 0.2719(7) c d$^{-1}$, or 3.678(9) d, 
the same period behind the large variations in light seen in Fig.~\ref{lc-ooe}.
Apparently the centre of 
light, or at least the centre of eclipsed light, wanders back and forth on this 
period.  And since the eclipsed light of UX UMa is dominated by the accretion 
disc, we conclude that the disc's photometric centre moves about with this 
period.\footnote{Where `disc' may or may not include the bright spot 
arising from mass transfer, which is a well-known permanent feature which causes 
the large asymmetry in the eclipse centred around orbital phase 0.05 
(see Fig.~9 of \citet{nather74}).}
(Presumably the true orbital period, set by the laws of dynamics, can be relied
on to stay immoveable during this 5-month campaign.)
A fold of the residuals on the ephemeris given in Eq.~(\ref{ephem_nodal}) 
yields the result 
seen in the lower frame of Fig.~\ref{p04-res}: a nearly sinusoidal wiggle with a 
semi-amplitude of 33.3(9) s. It seems that there is no time lag 
($O-C=0$) in the orbital eclipses at precession phases $\phi_p \approx 0$
and 0.5 (maximum and minimum light, respectively). The eclipse takes place
earlier than expected (negative residuals) for 
$0 < \phi_p < 0.5$, and later (positive residuals) for
$0.5 < \phi_p < 1$.

\subsection{The 0.1867 day (negative superhump) clock}

We tried to time individual maxima in the negative-superhump cycle 
by picking out local maxima after removing the 3.68 d and orbital signals.
Table~\ref{timings_nsh} shows the resultant 161 timings, with errors in 
individual timings estimated to be around 0.007 d. A linear regression
to these timings provides the ephemeris

\begin{equation}
T_{\rm{max}} ({\rm HJD}) = \;2,457,098.499(3) 
+ 0.186700(11) \,E  \, .
\label{ephem_nsh}
\end{equation}

\begin{figure}
\includegraphics[angle=-90,width=\columnwidth]{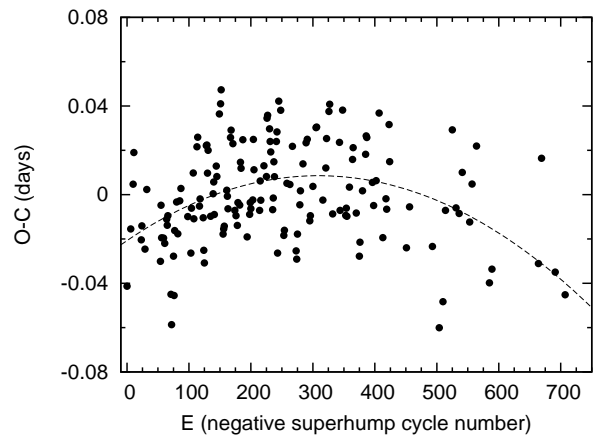}

\caption{
$O-C$ diagram of the 0.1867 d superhump maxima with respect to the test
ephemeris given in Eq.~(\ref{ephem_nsh}).}
\label{oc-nsh}
\end{figure}

The associated frequency, $\omega_{\rm nsh} = 5.3562(3)$ c d$^{-1}$, is fully
consistent with the value found from the power spectrum of the out-of-eclipse
photometric data.
The corresponding $O-C$ residuals relative to the above ephemeris are 
shown in Fig.~\ref{oc-nsh}. The downward curvature of the residuals
mirrors that of Fig.~\ref{oc-nodal}, verifying that the 
observed superhump frequency changes 
in lockstep with the observed precession frequency.  
From a parabolic fit to the residuals, we find that the period of the negative
superhump decreases at a rate 
$dP_{\rm nsh}/dt = -3.3(6) \times 10^{-6}$ or
$d\omega_{\rm nsh}/dt = 9.5(1.7) \times 10^{-5}$ c d$^{-2}$.
The latter agrees with the rate of variation of the nodal frequency we 
found before, as it should be expected from the
relation $\omega_{\rm nsh} = \omega_{\rm orb} + N$. We note that this relation
remains valid in the short term, not just for the whole season.  

\begin{table*}
 \centering
\caption{Times of maximum light on the 0.1867 day cycle
(${\rm HJD} - 2,457,000$).}
\centering
\begin{tabular}{rrrrrrrrrrr}
\hline
  98.458 &  99.604 & 100.371 & 100.572 & 102.773 & 102.966 & 103.889 & 
 104.476 & 108.551 & 108.763 & 108.935 \\
 109.495 & 109.866 & 110.437 & 110.621 & 110.812 & 111.710 & 111.883 &
 112.474 & 112.643 & 112.859 & 113.432 \\ 
 113.791 & 114.366 & 114.745 & 116.786 & 117.703 & 117.910 & 118.486 &
 118.652 & 119.618 & 119.809 & 120.338 \\
 120.528 & 121.453 & 121.625 & 121.806 & 122.419 & 122.606 & 122.780 &
 122.977 & 123.694 & 124.451 & 124.643 \\
 124.815 & 125.397 & 125.579 & 126.354 & 126.732 & 126.925 & 127.420 &
 127.609 & 127.797 & 128.560 & 128.744 \\
 128.925 & 129.704 & 129.894 & 130.448 & 130.978 & 131.349 & 131.718 &
 131.915 & 132.474 & 132.680 & 132.864 \\
 133.437 & 134.700 & 135.457 & 135.649 & 135.833 & 136.397 & 136.611 &
 136.784 & 138.446 & 138.646 & 138.824 \\
 139.773 & 140.515 & 140.728 & 140.916 & 141.470 & 141.651 & 141.833 &
 142.367 & 142.559 & 142.762 & 142.942 \\
 143.518 & 143.709 & 143.841 & 144.283 & 144.839 & 145.716 & 145.905 &
 146.673 & 147.606 & 148.370 & 149.443 \\
 149.626 & 149.824 & 150.584 & 150.777 & 151.536 & 152.479 & 152.854 &
 153.564 & 153.753 & 154.513 & 155.473 \\
 155.660 & 157.494 & 158.442 & 158.642 & 159.401 & 159.591 & 160.475 &
 162.561 & 162.717 & 163.509 & 164.395 \\
 164.585 & 164.768 & 165.528 & 166.474 & 166.666 & 167.570 & 168.484 &
 168.677 & 169.447 & 170.397 & 170.592 \\
 170.778 & 172.438 & 172.801 & 173.559 & 174.523 & 175.587 & 176.538 &
 176.720 & 177.505 & 177.675 & 182.677 \\
 183.629 & 190.519 & 192.536 & 193.668 & 194.456 & 196.546 & 197.631 &
 198.562 & 199.514 & 201.732 & 202.496 \\
 203.820 & 207.679 & 208.432 & 222.437 & 223.418 & 227.474 & 230.451 &
         &         &         &         \\
\hline
\label{timings_nsh}
\end{tabular}
\end{table*}

     The departures from a smooth curve are quite large 
-- up to 45 min --
whereas we 
estimate a typical measurement error of 10-15 min.  But the dispersion in timings 
on individual nights is much smaller, so we suspected that some other effect 
contributes to that variance. The power spectrum of the residuals about the
quadratic fit shows a peak at 0.273(2) c d$^{-1}$, 
which indicates that the precession 
term is responsible for this effect, even though its direct photometric 
signature -- the 3.68 d signal -- has been accurately subtracted.

 \section{Discussion}

      Most cataclysmic variables show a periodic signal at $P_{\rm orb}$, either 
from an eclipse -- pretty obvious! -- or from some other effect of high or moderate 
inclination, e.g. the periodic obscuration of the mass-transfer `hot spot' as 
it wheels around the disc.  Many ($\sim$200) also show a photometric period a few 
percent longer than $P_{\rm orb}$ (positive superhumps).  Most of the latter are 
short-period dwarf novae, which sprout these signals for 1--4 weeks, during 
their long outbursts (`supermaxima').  This is now understood as arising 
from the apsidal precession of the accretion disc, rendered eccentric at the 
3:1 resonance in the disc.  A few stars which are not dwarf novae also show this 
effect, but these are all short-period ($< 3.5$ h) nova-like variables, which 
in many ways can be seen as permanently erupting dwarf novae.  
These signals are known as `permanent' superhumps 
\citep{joep91_v603aql}.

      Only a disc large enough to reach the 3:1 resonance can suffer this 
instability \citep{whitehurst91,lubow91},
and that is presumably the reason that positive superhumps are only found in 
short-period stars. But some stars show photometric signals with 
$P < P_{\rm orb}$ 
-- the negative superhumpers.  Much less is known about them.  
The early papers on these phenomena 
\citep{bonnet-bidaud85,joep93_v603aql,joep95-v503cyg}
postulated the existence of a tilted accretion disc, which is forced to precess 
slowly backwards (relative to the orbit) by the torque from the secondary. 
The angular relation between the secondary (including its structures, viz. the 
mass-transfer stream) and the disc then repeats with a period slightly less than 
$P_{\rm orb}$. This is a negative superhump.  Roughly 20 CVs show negative 
superhumps (see Table~2 of \citet{montgomery09}), and roughly half of these 
(see Table~5 of \citet{armstrong13}) 
also show a photometric signal at the postulated 
precession period. Detection of that low-frequency signal is a strong point in 
support of the theory, since a wobbling disc should present an effective area 
which varies with the wobble period.

 Our data demonstrate that UX UMa joins this club.  We hypothesize that its 
accretion disc wobbles about the orbital plane with a period 
$P_{\rm nodal}=3.68$ d, and we 
see its effective area varying on that period.  But the orbiting secondary 
-- not  in the inertial frame! -- sees the disc with a slightly shorter recurrence 
period, such that 
$1/P_{\rm nsh} = 1/P_{\rm orb} + 1/P_{\rm nodal} = 5.356$ 
c d$^{-1}$, or $P_{\rm nsh} = 0.1867$ d.  
The effect is basically identical to the famous tropical/sidereal 
year effect in the Earth-Sun system, or the draconic/sidereal month effect in 
the Earth-Moon system. \citet{montgomery09} discusses this analogy in great, 
and fascinating, detail.

       The cause and maintenance of disc tilt is not known.  No actual dwarf 
nova in outburst shows negative superhumps, although their closest cousins 
-- nova-like variables with $P_{\rm orb} < 3.5$ h -- frequently do 
\citep{joep93_v603aql,armstrong13}.
It's possible that the 3:1 resonance is again involved, but with the tilt 
instability growing so slowly that only a `permanent' dwarf nova, which is in 
a high-viscosity state for a long time, can develop sufficient tilt.  An 
alternative theory is the recent work by 
\citet{thomas15},
which invokes white-dwarf magnetism to break the azimuthal symmetry and 
permit -- in fact, create -- disc tilt. 
They make an impressive case; and such an origin would be 
especially intriguing because UX UMa also shows the very-high-frequency DNOs 
(signatures of white-dwarf rotation?), which have remained equally mysterious.

 UX UMa is not a typical member of this club.  Most members belong to the
SW Sex subclass, which have shorter $P_{\rm orb}$ (3-4 h), occasional 
excursions to very low states, and only the $\omega_{\rm orb} + N$
feature (lacking $N$, and emphatically lacking $2\omega_{\rm orb} + N$).
They also commonly show periodic radial-velocity signals of high amplitude, 
presumably indicative of the mass-transfer stream overflowing the disc 
(because of the tilt).  Maybe CV zoology needs to be adjusted somewhat, 
in order to fit these oddities.

       Finally, why did we find all these new effects in a star which has been 
closely studied for 60 years?  Did they first arise in 2015?  It seems unlikely. 
Inspection of early light curves \citep{walker54,johnson54} reveals that both
the mean brightness and eclipse depths are not constant (see, for instance,
Table 1 in \citet{smak94}), with variations within the range we have observed 
in 2015. Also, \citet{knigge98b} 
have reported differences of up to 50 per cent in brightness
in {\it Hubble Space Telescope} 
({\it HST}) observations of UX UMa carried out 3 months apart
in 1994. They infer that a substantial  
($\sim 50$ per cent) variation of the mass transfer rate must have occurred, 
but a precessing disc during the 1994 {\it HST} observations would also account for
the observed brightness variations.
We therefore believe that the mean brightness and eclipse depths
were not exceptional in 2015.
We selected the star for observation partly because previously published light 
curves showed variations in the orbital waveform 
-- suggesting that a signal at 
some nearby frequency might be present.  But to actually reveal these effects, an 
extensive campaign is required, and no such campaign has ever been reported.  
So it's a decent bet, though by no means sure, that these superhump effects have 
been lurking, unsuspected, in many previous observations of UX UMa. 

 \section{Summary}

\begin{enumerate}
\renewcommand{\theenumi}{(\arabic{enumi})}

\item 

We report a long photometric campaign during 2015, with coverage on 
121 of 150 nights, totalling $\sim$ 1800 h. The star displayed a sinusoidal 
signal with a
semi-amplitude of 0.22 mag and a mean period of 3.680(7) d, or a 
frequency 0.2717(5) c d$^{-1}$. We identify the latter as $N$, the accretion 
disc's (putative) frequency of retrograde nodal precession.  

  \item 

Fig.~\ref{lcs-2nights} shows that the orbital waveform is highly variable from day 
to day, but not from orbit to orbit.  Power-spectrum analysis shows that 
this arises from signals non-commensurate with $P_{\rm orb}$, namely 
`negative superhumps' with frequencies
$\omega_{\rm orb}+ N$ and $2\omega_{\rm orb} + N$.

\item
 The mean orbital light curve -- shown in Fig.~\ref{lc-orb} 
and summed over more than 200 
orbits -- shows a wave with maximum light around orbital phase 0.35.  This is 
roughly 180$^{\circ}$ out of phase with the hot-spot effect seen in U Gem, which 
defines the standard accretion geometry for CVs.

\item
The 3.68 d period is strongly manifest in essentially every quantity we studied.  
The eclipse times wobble on this period with an amplitude of 33.3(9) s, probably 
because the disc's (projected) centre of light moves with that period.
The superhump times also wobble with that period, as do the eclipse depths. 

\item
Fig.~\ref{oc-nodal} shows that the precession period varied smoothly, decreasing 
by $\sim$0.2 d 
over the 5-month campaign. As it did, the superhump period changed accordingly,
maintaining $\omega_{\rm nsh} = \omega_{\rm orb} + N$.

\item
About a dozen other CVs show this basic triad of frequencies 
($\omega_{\rm orb}$, $N$, and $\omega_{\rm orb}+N$).  Most are so-called 
SW Sex stars. Because the physics which underlies this category is probably 
the wobbling non-coplanar disc, it is likely that the credentialing scheme of 
that club \citep{thorstensen91,rgil07,dhillon13} will have to change, in order to 
accommodate UX UMa.  We  note that \citet{neustroev11} has also, 
based on spectroscopic evidence, proposed that UX UMa has transient 
episodes of SW Sex behaviour.

\end{enumerate}

\section*{Acknowledgments}

We thank the National Science Foundation for support of this research 
(AST12-11129), and also the Mount Cuba Astronomical Foundation. Finally, we 
thank the American Association of Variable Star Observers (AAVSO) for providing 
the infrastructure and continued inspiration which makes programs like this 
possible.

\bsp

\label{lastpage}
\end{document}